\documentclass{ws-ijmpa}
\usepackage[super,compress]{cite}
\usepackage{graphicx}
\usepackage{url}
\usepackage{hyperref}
\usepackage[utf8]{inputenc}

\usepackage{soul}
\usepackage{xcolor}

\begin{document}

\newcommand{\BE}{\begin{equation}}
\newcommand{\EE}{\end{equation}}
\newcommand{\half}{{\scriptstyle{\frac{1}{2}}}}

\newcommand{\FF}[1]{{\color{blue}{}#1}}

\bibliographystyle{ws-ijmpa}

\markboth{M. Consoli, L.Cosmai, F. Fabbri}{Second resonance of the Higgs field: more signals from the LHC experiments}

\title{Second resonance of the Higgs field: more signals from the LHC experiments }

\author{Maurizio Consoli}
\address{INFN - Sezione di Catania,  I-95129 Catania, Italy \\
maurizio.consoli@ct.infn.it}

\author{Leonardo Cosmai}
\address{INFN - Sezione di Bari, I-70126 Bari, Italy\\
leonardo.cosmai@ba.infn.it}

\author{Fabrizio Fabbri}
\address{INFN - Sezione di Bologna, I-40127 Bologna, Italy\\
fabrizio.fabbri@bo.infn.it}

\maketitle

\pub{Received (Day Month Year)}{Revised (Day Month Year)}

\begin{abstract}
Theoretical arguments and lattice simulations suggest
that, beside the known resonance of mass $m_h=$ 125 GeV, the Higgs
field might exhibit a second resonance with a larger mass
$(M_H)^{\rm theor} = 690 \pm 10 ~({\rm stat}) \pm 20 ~({\rm sys})~
{\rm GeV}$ which, however, would couple to longitudinal W's with the
same typical strength as the low-mass state at 125 GeV and thus
represent a relatively narrow resonance mainly produced at LHC by
gluon-gluon fusion. By looking for some evidence in the LHC data, we argue 
that the existence of a new resonance in the predicted mass region finds support in two analyses by ATLAS (searching for heavy resonances decaying into
final states with 4 charged leptons or $\gamma\gamma$ pairs) and in 
more recent CMS results (searching for heavy resonances decaying
into a pair of $h(125)$ bosons or looking for $\gamma\gamma$ pairs
produced in $pp$ double-diffractive scattering). Since the correlation of these measurements is very small and since, having some definite theoretical prediction, local deviations from the pure background are
not downgraded by the look-elsewhere effect, we emphasize the instability of the present situation that could probably be resolved by just adding two crucial, missing samples of RUN2 data. 
\end{abstract}

\ccode{PACS numbers: 11.30.Qc; 12.15.-y; 13.85.-t}

\section{Introduction}

At present, the spectrum of the Higgs field is believed to consist
of just a single narrow resonance of mass $m_h=$ 125 GeV
defined by the second derivative of the effective potential at
its minimum. In a description of Spontaneous Symmetry Breaking (SSB)
as a second-order phase transition,  in the review of the Particle Data Group
\cite{Tanabashi:2018oca}, this point of view is well
summarized into a scalar potential of the form
\BE \label{VPDG} V_{\rm PDG}(\varphi)=-\frac{ 1}{2} m^2_{\rm PDG}
\varphi^2 + \frac{ 1}{4}\lambda_{\rm PDG}\varphi^4 \EE By fixing
$m_{\rm PDG}\sim$ 88.8 GeV and $\lambda_{\rm PDG}\sim 0.13$, this
has a minimum at $|\varphi|=\langle \Phi \rangle\sim$ 246 GeV and a
second derivative $V''_{\rm PDG}(\langle \Phi \rangle)\equiv m^2_h=$
(125 GeV)$^2$. Notice that, here, one is adopting the identification
$m^2_h= V''_{\rm PDG}(\langle \Phi \rangle)=|G^{-1}(p=0)|$ in terms
of the inverse, zero-momentum propagator.

However, lattice simulations of cutoff $\Phi^4$ in 4D
\cite{lundow2009critical,Lundow:2010en,akiyama2019phase} support
instead the view of SSB as a weak first-order phase transition.
While in the presence of gauge bosons SSB can indeed be described,
in perturbation theory,  as a first-order transition, recovering
this result in pure $\Phi^4$ requires to replace standard
perturbation theory with some alternative scheme. A scheme where SSB
occurs when the quanta of the symmetric phase have a very small but
still positive mass squared. Equivalently, a scheme where, as  in
the original Coleman-Weinberg (CW) 1-loop effective potential
\cite{Coleman:1973jx}, the massless, classically scale-invariant
theory is already found in the broken-symmetry phase. To exploit the
implications of this different description of SSB, a crucial
observation is that the CW potential admits two different readings.
In a first perspective, where the genuine 1-loop contribution is
understood as simply renormalizing the coupling $\lambda$ in the
classical potential,  the 1-loop minimum disappears after re-summing
the higher-order leading logarithmic terms.  In a second
perspective, recently emphasized in
\cite{Consoli:2020nwb,symmetry,memorial}, the same CW potential can
be read as the sum of the classical potential + the  zero-point
energy.  Once interpreted in this way, it also acquires a
non-perturbative meaning as the prototype \cite{Consoli:1999ni} of
an {\it infinite} number of approximations, the Gaussian
\cite{Stevenson:1985zy} and post-gaussian calculations
\cite{Stancu:1989sk,Cea:1996pe},
 which correspond to very different resummations
 of graphs but effectively reproduce the same basic structure of
$V_{\rm eff}(\varphi)$: some classical
background + zero-point energy of a particle with some  $\varphi-$dependent mass $M(\varphi)$.
In these approximations, by defining $m^2_h=V''_{\rm eff}(\varphi)
$ at the minimum, $M_H$ as the value of $M(\varphi)$ at the
minimum,  and introducing the ultraviolet cutoff $\Lambda_s$
one finds  \cite{Consoli:2020nwb,symmetry,memorial}  the following pattern of scales   ($L= \ln (\Lambda_s/M_H)$)
\BE \label{scale}    \lambda \sim {L}^{-1}~~~~~~~~~~~~ m^2_h \sim \langle \Phi\rangle^2 \cdot {L}^{-1}
 ~~ ~~~~~~~~ ~~~~~~~ M^2_H \sim L\cdot m^2_h = K^2 \langle \Phi\rangle^2 \EE
$K$ being a cutoff-independent constant.  Furthermore, the relation  $V_{\rm eff}(\langle \Phi \rangle)\sim -M^4_H$
supports the cutoff independence of $M_H$, and therefore of $ \langle \Phi\rangle$, because the ground state energy is a Renormalization
Group invariant quantity, see \cite{Consoli:2020nwb,symmetry,memorial}.

Notice that the two masses do not scale uniformly so that when  $\Lambda_s \to \infty$ one would be left
with only one mass, consistently with the free-field, continuum limit of the theory (``triviality''), see \cite{memorial}.
Still, the two masses $m_h$ and $M_H$ refer to
different momentum regions  in the propagator \footnote{The zero-point energy is
(one-half of) the trace of the logarithm of $G^{-1}(p)$. Therefore
$M_H$, reflecting also the behaviour of the propagator at large
Euclidean $p^2$, in general differs from $m^2_h\equiv
|G^{-1}(p=0)|$.} and could coexist in the cutoff
theory. As such, this two-mass structure was checked with
lattice simulations of the propagator \cite{Consoli:2020nwb}. Then, by computing
$m^2_h$ from the $p\to 0$ limit of $G(p)$ and $M^2_H$ from its
behaviour at higher $p^2$, the lattice data are consistent with the
scaling trend $M^2_H\sim L m^2_h $
and in the full momentum region can be described by a model form \cite{memorial}
\BE \label{interpol} G(p) \sim \frac{1 - I(p)}{2}\frac{1}{p^2 +
m^2_h }+\frac{1 + I(p)}{2}\frac{1}{p^2 + M^2_H } \EE
with an interpolating function $I(p)$ which depends on an
intermediate momentum scale $p_0$ and tends to $+1$ for large
$p^2\gg p^2_0$ and to $-1$ when $p^2 \to 0$. Therefore the lattice
data support the idea that, beside the known resonance with mass
$m_h=$ 125 GeV, the Higgs field might exhibit a second resonance
with a much larger mass. At the same time, by extrapolating from
various lattice sizes, the cutoff-independent constant was found
$K=2.80 \pm 0.04 ({\rm stat}) \pm 0.08 ({\rm sys})$ giving the
prediction \cite{Consoli:2020nwb,symmetry,memorial}\BE
\label{prediction} (M_H)^{\rm Theor} = 690 \pm 10 ~({\rm stat}) \pm
20 ~({\rm sys})~ {\rm GeV}\EE  The above numerical estimate, on the
one hand, represents a definite prediction to compare with
experiments.  On the other hand, it helps to clarify the relation
with the more conventional picture where there is only $m_h$. To
this end, we note that, from the third relation in (\ref{scale}),
one finds $m_h \ll M_H$ for very large $\Lambda_s$. But $M_H$ is
$\Lambda_s-$independent so that by decreasing $\Lambda_s$ also the
lower mass increases by approaching  its maximum value $(m_h)^{\rm
max}\sim M_H$  for $L\sim 1$, i.e. when the cutoff $\Lambda_s$ is a
few times $M_H$. Therefore this maximum value corresponds to \BE
 (m_h)^{\rm max}\sim (M_H)^{\rm Theor} =   690 \pm 10 ~({\rm stat}) \pm 20 ~({\rm sys})~
{\rm GeV}\EE in good agreement with the upper bound derived from the conventional first two relations  in Eq.(\ref{scale})
$(m_h)^{\rm max}= $ 670 (80) GeV, see Lang's complete review \cite{lang}. Equivalently,
without performing our own lattice simulations of the propagator, we could have predicted
$(M_H)^{\rm Theor}=$ 670 (80) GeV by combining the $\Lambda_s-$independence of $M_H$,
the third relation in (\ref{scale}) and the estimate of $(m_h)^{\rm max}$ in Lang's review paper.  This means
that for $\Lambda_s \to \infty$ the two masses decouple from each other while for
$\Lambda_s=O(1)$ TeV the two masses coincide. However, the physical situation has
$m_h=$ 125 GeV so that, if $M_H$ exists, $\Lambda_s$ would be very large.
At the same time, since vacuum stability would depend on the large $M_H$, and not on $m_h$,
SSB could originate within the pure scalar sector regardless
of the other parameters of the theory, e.g. the vector boson and top
quark mass.

After this preliminary introduction, in this paper we will first
briefly summarize in Sect.2 the expected phenomenology of the second
resonance and then, in Sects.3, 4 and 5, argue that the existence of a new resonance in the predicted mass range find support in two analyses by ATLAS (searching for heavy resonances decaying into
final states with 4 charged leptons  or $\gamma\gamma$ pairs) and in 
more recent CMS results (searching for heavy resonances decaying
into a pair of $h(125)$ bosons or looking for $\gamma\gamma$ pairs
produced in $pp$ double-diffractive scattering).  Finally, Sect.6 will contain a summary and our conclusions. 

\section{Phenomenology of the second resonance}

In spite of their substantial difference, the two masses $m_h$ and
$M_H$ describe excitations of the {\it same} Higgs field. Therefore
the observable interactions of this field, with itself and with the
other fluctuations about the minimum of the potential, as the
Goldstone bosons, are controlled by a single coupling: the mass
parameter $m^2_h$ determining the boundary condition at the Fermi
scale for the scalar coupling $\lambda = 3 m^2_h/ \langle \Phi
\rangle^2$. Thus, in spite of its large mass, the $H-$resonance
would couple to longitudinal W's and Z's with the same typical
strength as the low-mass state \footnote{We emphasize that the same
result is also recovered in a unitary-gauge calculation of
longitudinal W's scattering. There, at asymptotic energies and at
the tree-level, the $M^2_H$ in the Higgs propagator is promoted to
effective contact coupling $\lambda_0= 3M^2_H/\langle \Phi
\rangle^2$. But by re-summing higher-order terms with the
$\beta-$function, at the Fermi scale one finds $\lambda_0\to
\lambda=3m^2_h/\langle \Phi \rangle^2\sim L^{-1}$, see
\cite{memorial} and also \cite{Castorina:2007ng}.} at $125$ GeV and
represent a relatively narrow resonance. This is illustrated by the
replacement of the conventional large width for a heavy Higgs
particle $\Gamma^{\rm conv}(H \to WW+ZZ) \sim G_F M^3_H$ with the
corresponding relation $\Gamma(H \to WW+ZZ) \sim M_H (G_F m^2_h)$
which defines the same phase-space factor $M_H$ with a coupling
re-scaled by the small ratio $m^2_h/M^2_H\sim $ 0.032. Numerically,
for $M_H \sim $ 700 GeV, from the results of ref.\cite{handbook16},
this gives \footnote{As compared to the old values reported in \cite{handbook}, in the more recent ref.\cite{handbook16}, there has been a substantial reduction of the conventional $G_F M^3_H$ decay widths into W's and Z's, from 56.7 and 115.2 GeV to 50.1 and 102.6 GeV respectively.}\BE \label{rel1} \Gamma(H \to ZZ)\sim \frac{M_H}{700~{\rm
GeV}}\cdot\frac{m^2_h}{(700~{\rm GeV})^2}~50.1~{\rm GeV} \sim
\frac{M_H}{700~{\rm GeV}}\cdot ~1.6~ {\rm GeV}\EE \BE \label{rel2}
\Gamma( H \to WW)\sim \frac{M_H}{700~{\rm
GeV}}\cdot\frac{m^2_h}{(700~{\rm GeV})^2}~102.6~{\rm GeV} \sim
\frac{M_H}{700~{\rm GeV}}\cdot ~3.3~ {\rm GeV}\EE On the other hand,
the decays into fermions, gluons, photons..., which are proportional
to the gauge and yukawa couplings, would be unchanged and can be
taken from \cite{handbook16} yielding \BE \Gamma(H\to {\rm fermions+
gluons+ photons...})\sim \frac{M_H}{700~{\rm GeV}}\cdot 26.2~{\rm
GeV}\EE Therefore, one might expect a total width $\Gamma_H\equiv
\Gamma( H \to all) = 30\div 31$~{\rm GeV}. This estimate, however,
does not account for the new contributions from the decays of the
heavier resonance into the lower-mass state at 125 GeV. These
include the two-body decay $H\to hh$, the three-body processes $H\to
hhh$, $H\to hZZ$, $H\to hW^+W^-$ and the higher-multiplicity final
states allowed by phase space. For this reason, the above value
$30\div31$~{\rm GeV} should only be considered as a lower
bound.

It is not so simple to evaluate the new contributions to the total
decay width because of the $h-H$ overlapping which makes this a
non-perturbative problem. Perhaps, techniques as the RSE approach
\cite{RSE}, used in analogous problems with meson resonances, could
be useful. For this reason, in ref.\cite{signals}, we 
considered a test in the ``golden'' 4-lepton channel that does {\it
not} require the knowledge of the total width but only relies on two
assumptions:

~~ a) a resonant 4-lepton production through the chain $H \to ZZ \to
4l$

~~ b) the estimate of $\Gamma(H \to ZZ)$ in Eq.(\ref{rel1})

Therefore, by defining $\gamma_H=\Gamma_H/M_H$, we find a fraction
\BE B( H \to ZZ)= \frac {\Gamma( H \to ZZ)}{\Gamma_H}\sim \frac
{1}{\gamma_H}\cdot \frac{50.1}{700} \cdot\frac{m^2_h}{(700~{\rm
GeV})^2} \EE that will be replaced in the cross section approximated
by on-shell branching ratios \BE\label{exp3} \sigma_R (pp\to H\to
4l)\sim \sigma (pp\to H)\cdot B( H \to ZZ) \cdot 4 B^2(Z \to l^+l^-)
\EE This should be a good approximation for a relatively narrow
resonance so that one predicts a particular correlation
\BE\label{exp33} \gamma_H\cdot \sigma_R (pp\to H\to 4l)\sim \sigma
(pp\to H)\cdot \frac{50.1}{700}\cdot \frac{m^2_h}{(700~{\rm
GeV})^2}\cdot 4 B^2(Z \to l^+l^-) \EE which can be compared with the
LHC data.

Since $4B^2( Z \to l^+l^-)\sim 0.0045$, to check our picture, the
only missing piece is the production cross section $\sigma (pp\to
H)$ which, as discussed in \cite{memorial}, will mainly proceed
through the gluon-gluon Fusion (ggF) process. In fact, production
through Vector-Boson Fusion (VBF) plays no role once the large
coupling to longitudinal W's and Z's is suppressed by the small
factor $m^2_h/M^2_H\sim $ 0.032. Therefore, the sizeable VBF cross
section $\sigma^{\rm VBF}(pp\to H)\sim $ 300 fb is reduced to about
10 fb and is negligible with respect to the pure ggF contribution O($10^3$) fb. Indeed, for 13 TeV pp collisions, and with a typical $\pm 15\%$ uncertainty (due to the parton distributions, to the choice of $\mu$ in
$\alpha_s(\mu)$ and to other effects), from ref.\cite{yellow17} we find a value $\sigma^{\rm ggF} (pp\to H)= $ 1090(170) fb which also accounts for
the range $M_H=660\div 700$ GeV,  

In conclusion, for $m_h=$ 125 GeV,
we obtain a sharp prediction which, for not too large
$\gamma_H$ where Eq.(\ref{exp3}) looses validity, is formally
insensitive to the value of $\Gamma_H$ and can be compared with future high-precision
4-lepton data \footnote{The 12$\%$ reduction of the $\Gamma( H \to ZZ)$ decay width, from 56.7 to 50.1 GeV \cite{handbook16}, together with most recent estimate of the cross section $\sigma^{\rm ggF} (pp\to H)= $ 1090(170) fb, has produced a sizeable reduction with respect to the value 0.0137(21) fb of ref.\cite{signals}.}
\BE\label{exp34} [\gamma_H\cdot \sigma_R (pp\to H\to
4l)]^{\rm theor} \sim (0.011 \pm 0.002)~ {\rm fb} \EE

\section{The ATLAS  ggF-like 4-lepton events}

To obtain indications on a possible new resonance around 700 GeV, we started from the `golden' 4-lepton channel by considering the ATLAS sample \cite{ATLAS2,atlas4lHEPData} of events that, for their typical characteristics, admit the interpretation of being produced through the ggF mechanism. 

For these 4-lepton data, the ATLAS experiment has performed a multivariate
analysis (MVA)  which combines a multilayer perceptron (MLP) and one or two recurrent neural networks (rNN). The outputs of the MLP and rNN(s) are concatenated so as to produce an event score. In this way, depending on the score, the ggF events are divided into four mutually exclusive categories:
ggF-MVA-high-4$\mu$, ggF-MVA-high- 2e2$\mu$, ggF-MVA-high-4e, ggF-MVA-low. 

This class of ggF-like events was already considered in ref.\cite{signals} 
with the conclusion that there are definite indications for a new resonance in the expected mass region. However, due to some model-dependent assumptions in the analysis of the data, here we have decided to adopt a different strategy. The starting point was to consider the other ATLAS article on the differential cross section, as function of the 4-lepton invariant mass \cite{atlasnew}, and in particular their Fig.5. The figure indicates a sizeable excess of events around 680 GeV which, soon after,is followed by a corresponding sizeable defect. Notice that the bins used at high masses have size of 60 GeV or more, probably to minimize smearing effects due to the invariant mass resolutions of the different final states. Indeed, for a 700 GeV resonance, the resolutions in invariant mass for $4e$, $2e2\mu$ and $4\mu$ final states are approximately 12 GeV, 19 GeV and 24 GeV respectively \cite{atlasresolution} so that a bin of 60 GeV, centered at a given value with ± 30 GeV range, is large enough not to be significantly affected by smearing effects (with spurious migrations of events between neighboring bins). 

By expecting our second resonance to be produced through gluon-gluon fusion (ggF) we then looked for indications in that particular sector and considered the category of the ggF-low events, which are homogeneous from the point of view of the selection and have sufficient statistics. At the same time, since this category contains a mixture of all three final states, it is natural to follow the above large-bin strategy. In our understanding, the ggF-low sample is certainly less pure as compared to the ggF-high samples, and it is true that it includes the dominating contribution from non-resonant ZZ events. On the other hand, according to ATLAS, this total background was carefully evaluated with a quoted total (stat. + syst.) uncertainty which is rather small (less than $5\%$ in the relevant region \cite{atlas4lHEPData}). As such, we see no reason not to consider it as our best estimate and safely subtract the background from the observed events. The result of this subtraction is shown in Table 1.

\begin{table}[htb]
\tbl{For luminosity 139 fb$^{-1}$, we report the observed ATLAS ggF-low events and the corresponding estimated background \cite{atlas4lHEPData} in the range of invariant mass $M_{4l}=E=530\div 830$ GeV. To avoid spurious fluctuations, due to migration of events between neighbouring bins, we have followed the same criterion as in Fig.5 of ref.\cite{atlasnew} by grouping the data into larger bins of 60 GeV, centered at 560, 620, 680, 740 and 800 GeV. These were obtained by combining the corresponding 10 bins of 30 GeV, centered respectively at the neighbouring pairs: 545(15)$\div$ 575(15) GeV, 605(15)$\div$ 635(15) GeV, 665(15) $\div$ 695(15), 725(15)$\div$755(15) GeV and 785(15)$\div$ 815(15) GeV as reported in ref.\cite{atlas4lHEPData}. In this energy range, the errors in the background are below $5\%$ and will be ignored.}
{\begin{tabular}{@{}cccc@{}} 
\toprule
$\rm E$[GeV] & $\rm{N}_{\rm EXP}(E)$  & $\rm{N}_{\rm bkg}(E)$ & $\rm{N}_{\rm EXP}(E) - \rm{N}_{\rm bkg}(E)$ \\
\hline
560(30) & 38$\pm 6.16$ & 32.0 & $ 6.00 \pm 6.16$  \\
\hline
620(30) & 25$\pm 5.00$ & 20.0 & $5.00 \pm 5.00$  \\
\hline
680(30) & 26$\pm 5.10$ & 13.04 & $12.96 \pm 5.10$ \\
\hline
740(30) & 3$\pm 1.73$ & 8.71& $-5.71 \pm 1.73$  \\
\hline
800(30) & 7$\pm 2.64$ & 5.97 & $1.03 \pm 2.64$  \\
\botrule
\end{tabular}}
\end{table}

From our Table 1, one gets the same qualitative impression as from Fig.5 of ref.\cite{atlasnew} in the same energy region. Namely, a sizeable (2.5-sigma) excess of events over the background, in the bin centered around 680 GeV, followed by a sizeable opposite defect (about 3-sigma) in the bin centered around 740 GeV. The simplest explanation for these two simultaneous features would be the existence of a resonance of mass $M_H\sim$ 700 GeV which, beside the resonant peak, by interfering with the non-resonating background produces the characteristic  change of sign of the interference term proportional to $(M^2_H-s)$. 

We have thus attempted to describe the data in Table 1 by using 
the same model cross section adopted in \cite{memorial,signals}
\BE \label{sigmat} \sigma_T=\sigma_B
+\frac{2(M^2_H -s)~\Gamma_H M_H}{(s-M^2_H)^2+ (\Gamma_H
M_H)^2}~\sqrt{\sigma_B\sigma_R} +\frac{(\Gamma_H M_H)^2
}{(s-M^2_H)^2+ (\Gamma_H M_H)^2}~ \sigma_R\EE 
which, together with the mass $M_H$ and total width $\Gamma_H$ of the resonance,  introduces a background cross section $\sigma_B=\sigma_B(E)$ and a peak cross section $\sigma_R$. Of course, due to the very large size of the bins, ours should only be considered a first, rough approximation. Nevertheless, we found a good consistency with the phenomenological picture of Sect.2. 

\begin{figure}[ht]
\centering
\includegraphics[width=0.5\textwidth,clip]{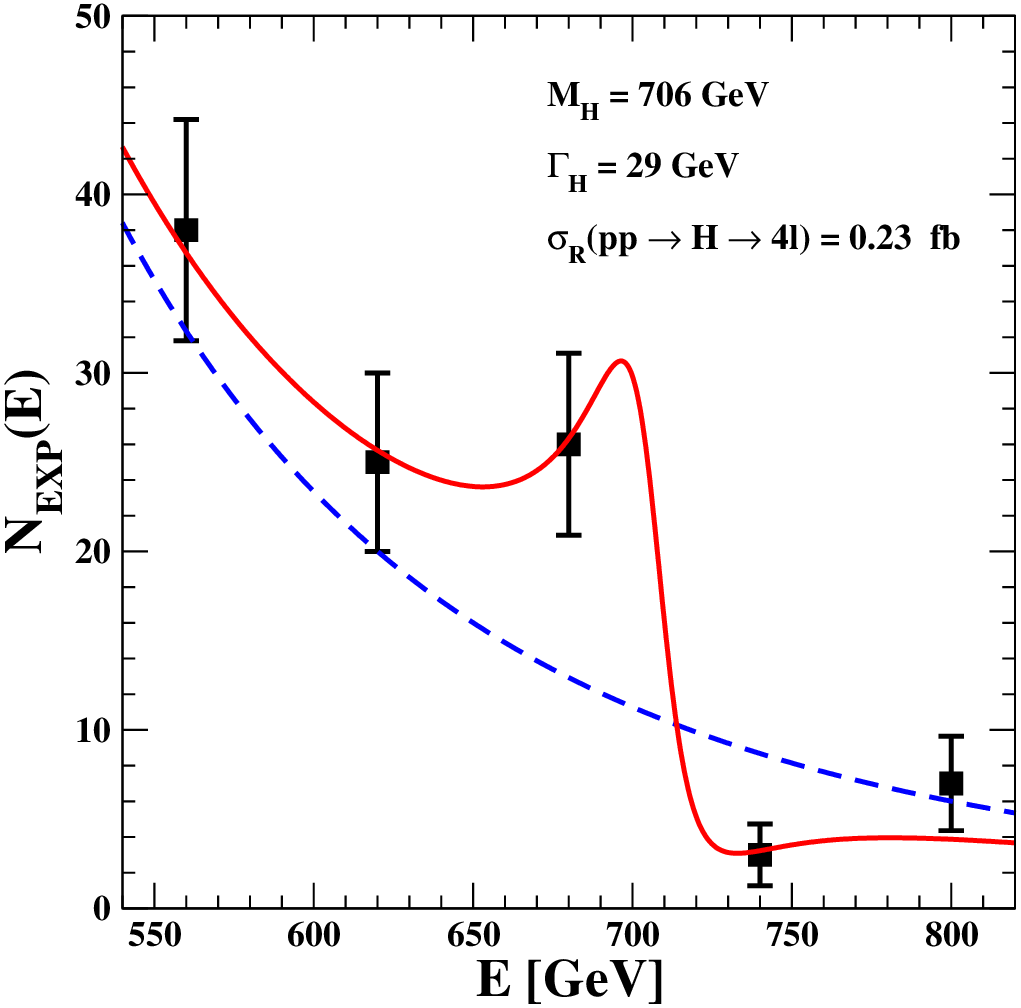}
\caption{The observed ATLAS ggF-low events are compared with Eq.(\ref{NTH}) (the red continuous line) for resonance parameters $M_H=$ 706 GeV, $\gamma_H=$0.041 and peak cross section $\sigma_R=$ 0.23 fb (equivalent to $P=0.14$ and $N_R\sim$ 12). The dashed blue line models ATLAS background as
$N_B(E)\sim A\cdot({\rm 710~GeV}/E) ^{\nu}$ with $A\sim 10.55$ and $\nu\sim 4.72$.} \label{4lepton}
\end{figure}

For our comparison, we first searched for an accurate description of the ATLAS background in terms of a power law $N_B(E)\sim A\cdot({\rm 710~GeV}/E) ^{\nu}$ with $A\sim 10.55$ and $\nu\sim 4.72$. Then, by simple redefinitions, the theoretical number of events can be  expressed as 
\BE \label{NTH} N_{TH}(E)= N_B(E) + \frac{P^2 + 2 P\cdot x(E)\cdot \sqrt{N_B(E)} }{\gamma^2_H + x^2(E)}~ \EE
where $x(E)=(M^2_H -E^2)/M^2_H$, and $P$ is defined as $P \equiv \gamma_H\sqrt{N_R} $ in terms of the extra number of events at the resonance peak $N_R=\sigma_R \cdot {\cal A }\cdot 139$ fb$^{-1}$ for given acceptance ${\cal A }$ and luminosity. Without a specific information on the acceptance of the ggF-low category, we adopted a value ${\cal A }\sim 0.38$ by averaging the two extremes, 0.30 and 0.46, quoted for the lowest and highest mass regions \cite{ATLAS2}. As a consequence, the resonance parameters will be affected by a corresponding uncertainty. We then fitted with Eq.(\ref{NTH}) the experimental number of events in Table 1. The results were: $M_H=$ 706(25) GeV, $\gamma_H= 0.041\pm 0.029$ (corresponding to a total width $\Gamma_H= 29\pm 20$ GeV) and $P= 0.14 \pm 0.07$. From these we obtain central values $\langle N_R \rangle \sim $12 and $\langle \sigma_R\rangle \sim$ 0.23 fb with very large errors. Our theoretical values are shown in Table 2 and a graphical comparison in Fig.\ref{4lepton}.

\begin{table}[htb]
\tbl{The experimental ATLAS ggF-low events are compared with our theoretical prediction Eq.(\ref{NTH}) for $M_H=$ 706 GeV, $\gamma_H=$ 0.041, $P=$ 0.14.}
{\begin{tabular}{@{}cccc@{}} 
\toprule
$\rm E$[GeV] & $\rm{N}_{\rm EXP}(E)$  & $\rm{N}_{\rm TH}(E)$ & $\chi^2$ \\
\hline
560(30) & 38$\pm 6.16$ & 36.72 & 0.04  \\
\hline
620(30) & 25$\pm 5.00$ & 25.66 & 0.02  \\
\hline
680(30) & 26$\pm 5.10$ & 26.32 & 0.00 \\
\hline
740(30) & 3$\pm 1.73$ & 3.23& 0.02  \\
\hline
800(30) & 7$\pm 2.64$ & 3.87 & 1.40  \\
\botrule
\end{tabular}}
\end{table}
 
The quality of our fit is good but, with the exception of the mass, errors are very large and the test of our picture is not so stringent. Still, with the partial width of Sect.2, $\Gamma(H\to ZZ)\sim$ 1.6 GeV, and fixing $\Gamma_H$ to its central value of 29 GeV, we find a branching ratio $B(H\to ZZ)\sim$ 0.055 which, for the central value $\sigma^{\rm ggF} (pp\to H)\sim $ 923 fb \cite{handbook16} at $M_H=$ 700 GeV, would imply a theoretical peak cross section $(\sigma_R)^{\rm theor}= (923\cdot 0.055\cdot 0.0045)\sim $ 0.23 fb which coincides with the central value from our fit. Also, from the central values of our fits $\langle \sigma_R\rangle =$ 0.23 fb and $\langle\gamma_H\rangle =$ 0.041 we find $\langle \sigma_R\rangle \cdot \langle\gamma_H\rangle \sim$ 0.0093 fb consistently with the theoretical prediction Eq.(\ref{exp34}). 

In conclusion, the ATLAS ggF-low category of 4-lepton events indicate the existence of a new resonance whose mass and basic parameters are consistent with our picture.

\section{The ATLAS high-mass $\gamma \gamma$ events}

Looking for further signals, we have then
considered the ATLAS $\gamma\gamma$ events in the range of invariant
mass $600\div770$ GeV which extends about $\pm 90$ GeV around our
central mass value. The relevant entries in Table 3 were extracted
from Fig.3 of \cite{atlas2gamma} because the numerical values are
not reported in the companion HEPData file.

\begin{figure}[ht]
\centering
\includegraphics[width=0.5\textwidth,clip]{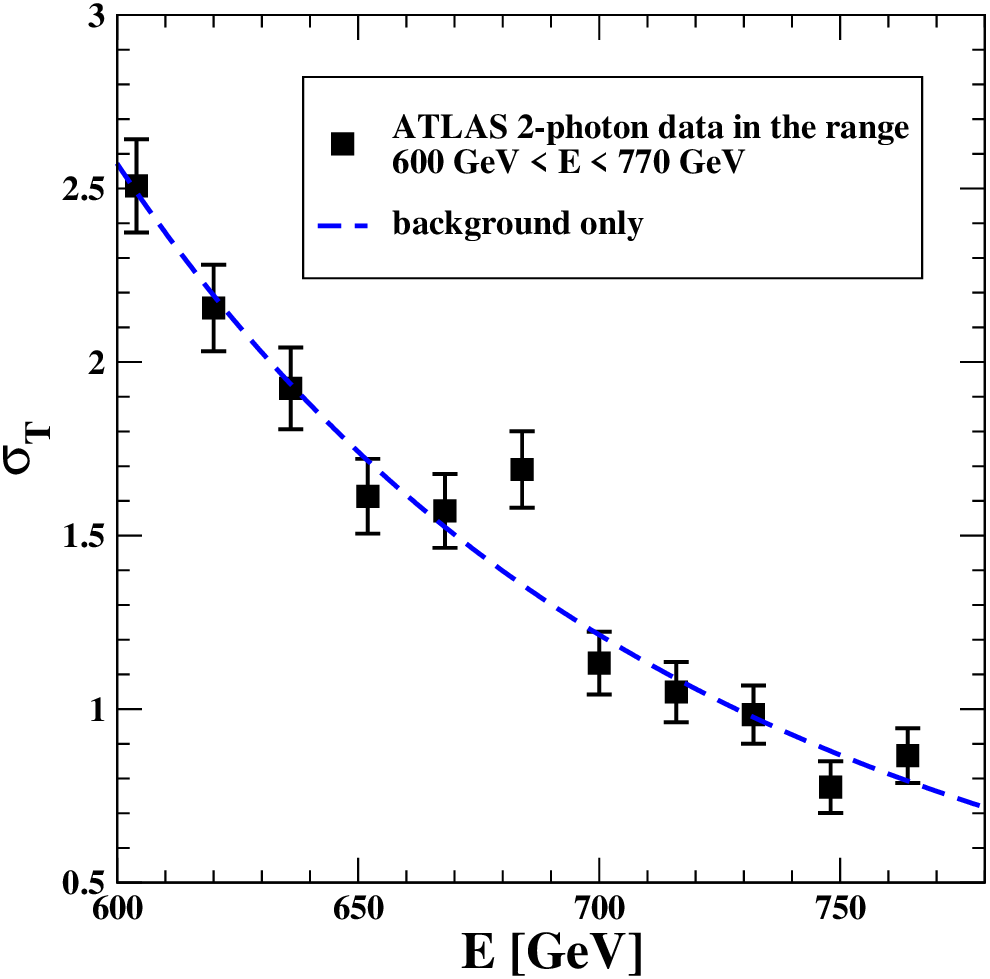}
\caption{The fit with Eq.(\ref{sigmat}) and $\sigma_R=0$ to the data
in Table 2, transformed into cross-sections in fb. The chi-square
value is $\chi^2=$ 14 and the background parameters $A=1.35$ fb and
$\nu=4.87$. } \label{twogamma0}
\end{figure}

Again, we have performed various fits to the
corresponding cross sections by parameterizing the background with a
power-law form $\sigma_B(E) \sim A\cdot ({\rm 685~GeV}/E) ^{\nu}$. This
gives a good description of all data, with the exception of
the sizeable excess at 684 GeV (estimated by ATLAS to have a local
significance of about 3.3-sigma). To have an idea, by fixing
$\sigma_R=0$ in Eq.(\ref{sigmat}), a pure background fit gives
$A=1.35(3)$ fb and $\nu=4.87(38)$ with $\chi^2=14$, but 10 of which
are only due to the peak at 684 GeV, see Fig.\ref{twogamma0}.

Therefore the (hypothetical) new resonance might
remain hidden by the large background almost everywhere, the
main signal being just the interference effect. However, since this interference effect is much smaller than the background in the resonance region ($\sigma_B \sim$ 1.3 fb), the resonance parameters will be
determined very poorly. In this perspective, our scope is not to
provide a better-quality fit but to show that the same
$\gamma\gamma$ data are also well consistent with the existence of a new
resonance in the expected mass region. 

\begin{figure}[ht]
\centering
\includegraphics[width=0.5\textwidth,clip]{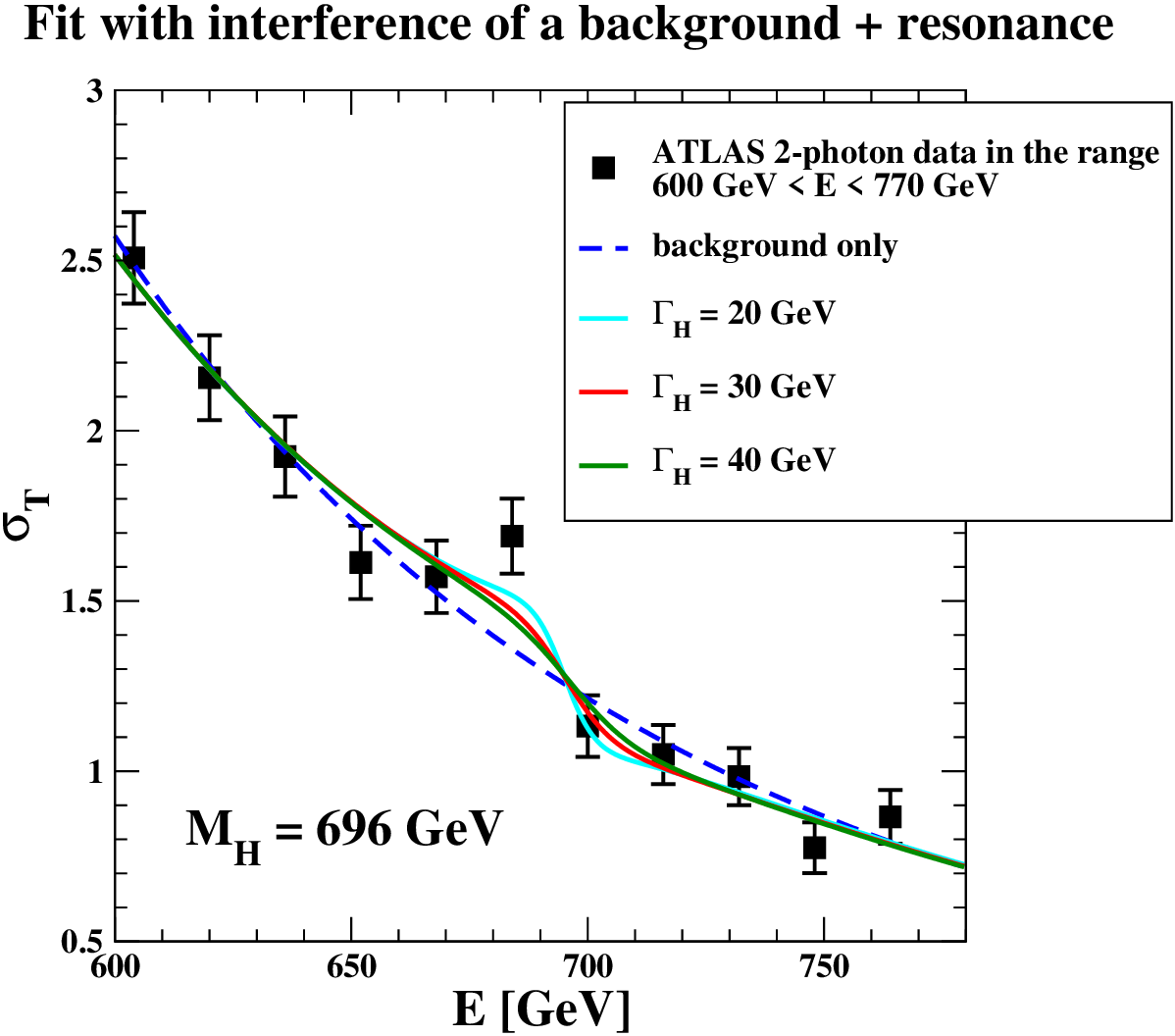}
\caption{Three fits with Eq.(\ref{sigmat}) to the data in Table 2,
transformed into cross-sections in fb. The $\chi^2-$values are 8.0,
9.5, 10.7 respectively for $\Gamma_H=$ 20, 30 and 40 GeV. }
\label{twogamma1}
\end{figure}

To this end, by constraining the background parameters $A$ and $\nu$
in $\sigma_B$ to lie within the region from the previous fit for
$\sigma_R=0$, we have performed several fits with the full
Eq.(\ref{sigmat}). Three fits are shown in Fig.\ref{twogamma1} for
$M_H=$ 696 GeV and for the values $\Gamma_H=$ 20, 30 and 40 GeV.

\begin{table}[htb]
\tbl{We report the ATLAS number of $N=N(\gamma\gamma)$ events,
in bins of 16 GeV and for luminosity 139 fb$^{-1}$, for the range of
invariant mass $\mu=\mu(\gamma\gamma)=600\div 770$ GeV. These entries
were directly extracted from Fig.3 of \cite{atlas2gamma} because the
relevant numbers are not reported in the corresponding HEPData
file.}
{\begin{tabular}{@{}cccccccccccc@{}} 
$\mu$&604&620&636&652&668&684&700&716&732&748&764 \\
\hline
$N$&349(19)&300(17)& 267(16)& 224(15)& 218(15)& 235(15)&
157(13)&146(12)&137(12)&108(10)& 120(11)\\
\botrule
\end{tabular}}
\end{table}

\begin{figure}[ht]
\centering
\includegraphics[width=0.5\textwidth,clip]{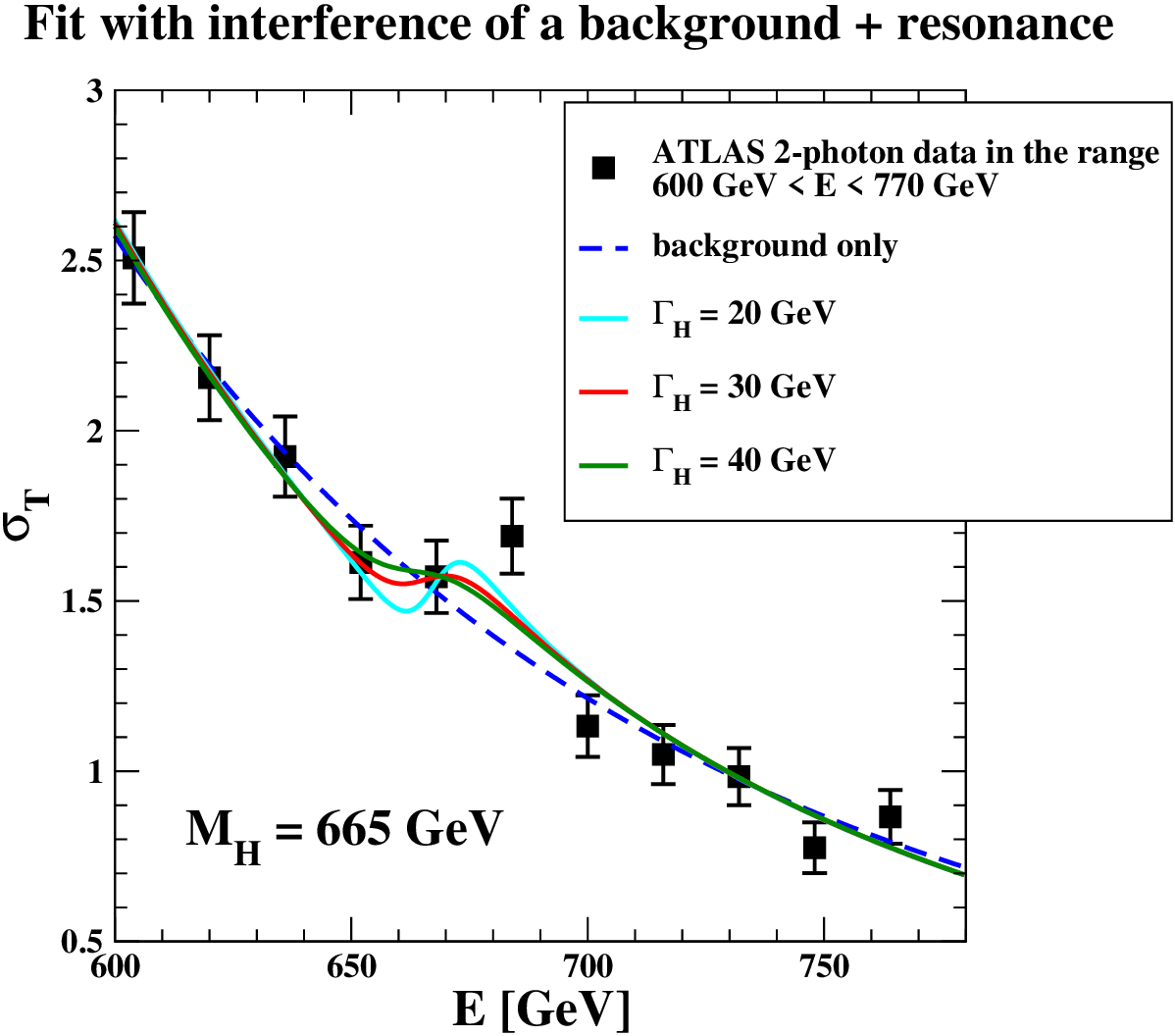}
\caption{Three fits with Eq.(\ref{sigmat}) to the data in Table 2,
transformed into cross-sections in fb. With respect to
Fig.\ref{twogamma1}, we have reversed the sign of the interference
term in Eq.(\ref{sigmat}). The
$\chi^2-$values are 9.6, 10.4 and 11.0 respectively for $\Gamma_H=$ 20, 30 and 40 GeV.} \label{twogamma2}
\end{figure}

Due to the predominant role of the interference effect, in
this $\gamma\gamma$ case, we also did a second series of fits by
reversing the sign of the interference term (from
positive to negative below peak) which is not known a
priori. For the same values of the width reported in
Fig.\ref{twogamma1}, the corresponding fits are shown in
Fig.\ref{twogamma2}. Note that, in this second type of fits there is
a shift of about $-$30 GeV, in the central value of the mass, from
$M_H =$ 696(12) GeV down to $M_H=$ 665(13) GeV. 

Thus, in principle, the localized 3.3-sigma excess at 684 GeV admits two different interpretations:

~~a) a statistical fluctuation above a pure background, see fit in
Fig.\ref{twogamma0} 

~~b) the signal of a new resonance, see 
Figs.\ref{twogamma1} and \ref{twogamma2} 

As for the total width, the $\gamma\gamma$ data tend to place
an upper limit which could be summarized as $(\Gamma_H)^{\rm exp} < 50$
 GeV. 
The analogous combined determination for the mass could
be summarized into $(M_H)^{\rm exp}=$ 680(15) GeV which summarizes
the results obtained with the two possible signs of the interference
term in Eq.(\ref{sigmat}).

As for the peak cross section  $\sigma_R=\sigma_R(pp\to H\to
\gamma\gamma)$, its values are always very small. The central values
lie in the range $0.004\div 0.009$ fb with minus errors, however,
which, at the 70$\%$ confidence level, extend very close to $\sigma_R \sim 0$
(where $\chi^2=14$). Therefore, before concluding this section, it
is appropriate to comment on the theoretical prediction for this
quantity which depends on the partial width $\Gamma(H \to
\gamma\gamma)$.

For the definite value $M_H = 700$ GeV, the estimate of
ref.\cite{handbook16}  is $\Gamma(H \to \gamma\gamma)=$ 29 keV.
Therefore, for $(\Gamma_H)^{\rm Exp} = 30\div 40$ GeV, one might expect
a branching ratio $B(H \to \gamma\gamma)\sim 8\cdot 10^{-7}$ and a
peak cross section $\sigma_R(pp\to H\to \gamma\gamma)\sim
\sigma(pp\to H)\cdot B(H \to \gamma\gamma)\sim 8 \cdot 10^{-4}$ fb.
However, it should be emphasized that this estimate of $\Gamma(H \to
\gamma\gamma)$ contains the non-decoupling, so called, ``$-2$'' term
proportional to $M^3_H$ whose existence (or not)in the WW
contribution has been discussed at length in the literature. At
present, the general consensus is that this term has to be there
with the only exception of the unitary-gauge calculations of
Gastmans, Wu and Wu \cite{GWuWu,WuWu,WuWu2} and the
dispersion-relation approach of Christova and Todorov
\cite{Todorov}. We believe that, in the context of a second Higgs
resonance, which does {\it not} couple to longitudinal W's
proportionally to its mass but with the same typical strength as the
low-mass state at 125 GeV, the whole issue could be re-considered.
The point is that for the range of mass around $M_H=$ 700 GeV there
are strong cancelations between the WW and $t \bar t$ contributions.
As we have checked, the presence or not of the non-decoupling term
could easily change the lowest-order value (i.e. without QCD
corrections in the top-quark graphs) by an order of magnitude. For
this reason, an increase of the partial decay width up to $\Gamma(H
\to\gamma\gamma) \lesssim$ 100 keV cannot be excluded. This would
increase the theoretical branching ratio and, therefore, the
theoretical peak cross section toward the region $\sigma_R(pp\to
H\to \gamma\gamma)= O(10^{-3})$ fb favoured by our fit to the
$\gamma\gamma$ events.

Of course, more precise data are needed for a significative test.
However since, most probably, even with the whole integrated
luminosity from RUN3, the statistical errors will not change
substantially, one could search for additional information from
other channels. For instance, from $\gamma\gamma$ exclusive
production in $pp$ double-diffractive scattering, i.e. in the process \BE
p+p\to p + X + p \EE when both final protons are tagged and have large Feynman $x_F$.
For $X=H$, which is the ``diffractive excitation of the vacuum''
considered by Albrow \cite{albrow}, one could then look for an
excess of $\gamma\gamma$ events in the region of invariant mass
$\mu(\gamma\gamma) \sim M_H$. Since, here, the background is
different, the corresponding excess could have a larger statistical
significance. Remarkably, as we will briefly discuss in the
following section, such an excess is indeed present in the CMS data.

\section{Experimental signals from CMS: a survey}

Unfortunately CMS analyses of charged 4-lepton and fully inclusive photon
pair events, with the whole statistics collected during RUN2,
are not yet available. However, we report below some indications from smaller statistical samples to get, at least, the qualitative indication of some excess in the relevant mass region. 

In this respect, the CMS 35.9 fb$^{-1}$ 4-lepton data show a clean excess around 660 GeV (see Fig.3, left panel of \cite{CMSJHEP17}), consistently with the corresponding ATLAS sample with the full 139 fb$^{-1}$
luminosity. This is also visible in first panel (up left) of Fig.2 of \cite{untagged} with the same clean statistical significance.  

The corresponding $\gamma\gamma$ data, for the same
statistics of 35.9 fb$^{-1}$, also show a modest 1-sigma excess for
the EBEB selection mode, see Fig.2, left panel of \cite{CMSPRD18}.
The data were grouped into bins of 20 GeV and the relevant concerns the three bins from 620(10) to 660(10) GeV. The analogous EBEE
plot also shows a very slight excess at 660 GeV but this is really too
small. The 1-sigma excess for 640(30) GeV in the EBEB is also visible in panel 5 of Fig.4 of \cite{CMSPRD18} where the data are compared with the case of
a scalar resonance, for $\gamma_H=$0.056 which is close to our model. Notice that the excess is followed by a corresponding defect of events as in the ATLAS 4-lepton case we have discussed, The corresponding plot in panel 6 for the EBEE mode shows instead only a defect of events.

While, to present date, there is no CMS analysis of high-mass $\gamma\gamma$
pairs with a statistics larger than 35.9 fb$^{-1}$, 
high-mass, 4-lepton events with the larger 77.4 fb$^{-1}$ luminosity were reported in \cite{CMS4lepton77}. Though, it is very hard to deduce anything from the very compressed scale adopted for the figures (see however Cea's analysis \cite{cea} claiming for a substantial excess around 700 GeV). From this point of view, these partial CMS measurements certainly do not contradict our present analysis.

But a second resonance of the Higgs field, if there, should also be visible in other decay channels as well. For this reason, we will now discuss some results presented by CMS at the ICHEP 2022 Conference and/or made recently publicly accessible. A first analysis  \cite{CMS PAS HIG-20-016}  concerns the search for high mass resonances decaying into  W-pairs which then decay into neutrinos and the charged leptonic final state ($e\mu$, $\mu\mu$, $ee$). The excesses observed have non-negligible local significances which range from 2.6 to 3.8 sigmas, depending on the different mechanisms for the resonance production mode (ggF and/or VBF). The scenario with VBF = 0 has the maximal significance for a resonance of mass 950 GeV, however, the region where one can see a deviation above 1 sigma is quite broad, 600 - 1200 GeV, due to the presence of neutrinos in the final state, and this is consistent with an excess in a mass region that also includes the one we predicted.

Another analysis  \cite{CMS PAS HIG-21-011}  concerns
the search for new resonances decaying, through two intermediate
scalars, into the peculiar final state  made by a $ b \bar b$ quark
pair and a $\gamma\gamma$ pair. In particular, here one has been
considering  the cross section for the full process \BE\sigma({\rm
full}) = \sigma (pp\to H \to hh \to b \bar b + \gamma\gamma)\EE (in
the CMS paper the new heavy resonance is called $X$ and the 125 GeV
resonance is called H while here we denote $X\equiv H$ and define
$h=h(125)$). For a spin-zero resonance, the 95$\%$ upper limit
$\sigma({\rm full})<0.16 $ fb, for invariant mass of 600 GeV, was
found to increase by about a factor of two, up to  $ \sigma({\rm
full})<0.30$ fb in a plateau $650\div 700$ GeV, and then to decrease
for larger energies. The local statistical significance is modest,
about 1.6-sigma, but the relevant mass region 675(25) GeV is precise
and agrees well with our analysis of the ATLAS data. Interestingly,
if the cross section is approximated  as \BE \sigma({\rm full})\sim
\sigma(gg\to H) \cdot B(H\to hh) \cdot 2 \cdot B(h \to b\bar b)B(h
\to \gamma \gamma) \EE after replacing our reference value
$\sigma(gg\to H) = 1090(170) $ fb, $ B(h \to b\bar b)\sim $0.57 and
$B(h \to \gamma \gamma)\sim$ 0.002, the CMS 95$\%$ upper bound
$\sigma({\rm full}) <  0.30$ fb gives a rather precise upper bound $
B(H\to hh) < $ 0.12. In view of the mentioned non-perturbative
nature of the decay process $H \to hh$ this represents a precious
indication.

Finally, as anticipated, CMS has been searching for high-mass
photon-pairs exclusively produced in $pp$ diffractive scattering,
i.e. when both final protons have large $x_F$. For our scopes, the
relevant information is contained in Fig.5, fourth panel (second
row, right)  of \cite{CMS PAS EXO-21-007}. In the range of invariant
mass 650(40) GeV, and for a statistics of 102.7 fb$^{-1}$ the observed number of $\gamma\gamma$ events was $N_{\rm obs}\sim$ 76(9) to be compared with an estimated background $N_{\rm BKG}\sim$ 40(6). In the most conservative case, this is a local excess of 3.3-sigma significance. 

Now, our understanding is that the uncertainty on the background of each bin has been evaluated by means of high-statistics samples which account for the most important processes ($t \bar t +$jet; $Z+\gamma$, $W+ \gamma$; $\gamma +$ jet; QCD events) fully simulated in the detector, reconstructed as the
real data and subjected to the same type of analysis as the real
data. For the relevant bin with invariant mass 650(40) GeV, this
sophisticated procedure gives a background estimate of 40 events
with an uncertainty of $\pm 6$ described by the hatched area in the
plot of Fig.5, fourth panel (second row, right) of \cite{CMS PAS
EXO-21-007}, see also the lower part for the ratio Data/Prediction.
As one can see, apart from two points with large errors around 1
TeV, all bins lie within the hatched area, {\it except} the bin at
650(40)GeV where the 76(9) observed events correspond indeed to the
mentioned 3.3-sigma excess. However, a critical observer cannot help
but notice that interpolating the background yields from the two
neighbouring bins would suggest a background estimate which is closer to
60 rather than 40. Starting from this remark, within the CMS
Collaboration, an effort should be made to see whether, in the end, 
with a more refined evaluation of the background (and adding the remaining  statistics of about 35 fb$^{-1}$ events) the excess will maintain the present, sizeable statistical significance. 

At present, the two excesses found in the two previous CMS analyses
point to a new resonance of mass  $(M_H)^{\rm Exp}\sim$ 670(20) GeV.

\section{Summary and conclusions}

In this paper, we started from the idea
\cite{Consoli:2020nwb,symmetry,memorial} that, beside the known
resonance with mass $m_h=$125 GeV, the Higgs field could exhibit a
second resonance with a much larger mass $M_H$. This new state,
however, would couple to longitudinal W's with the same typical
strength as the low-mass state at 125 GeV and thus represent a
relatively narrow resonance mainly produced at LHC by gluon-gluon
fusion (ggF). From theoretical arguments and lattice simulations,
its mass can be estimated to have a value $(M_H)^{\rm Theor} = 690
\pm 10 ~({\rm stat}) \pm 20 ~({\rm sys})$ GeV.

To find some signal in the LHC data, we started in Sect.3 with the ATLAS search for a new resonance in the charged 4-lepton  channel by considering the ggF-low category of events which forms a homogeneous sample and has sufficient statistics. As reported in Table 1, this sample shows a 2.5-sigma excess at 680(30) GeV followed by an opposite 3-sigma defect at 740(30) GeV. This feature is also confirmed by Fig.5 of \cite{atlasnew}. As we have argued, the simplest explanation would be the existence of a new resonance with mass $M_H\sim $ 700 GeV which, besides the Breit-Wigner peak, produces the characteristic $(M^2_H -s)$ interference effect. As shown in Table 2 and in Fig.\ref{4lepton}, this interpretation is consistent with the data for the range of parameters suggested by the phenomenological picture of Sect.2. 

After this first indication, we have considered in Sect.4 the ATLAS $\gamma\gamma$ events in the high-mass region $600\div770$ GeV. As we have shown, see Figs.\ref{twogamma0}-\ref{twogamma2}, the sizeable 3.3-sigma (local) excess at 684 GeV in the $\gamma\gamma$ distribution can also be
interpreted as the interference signal, with a dominating background, of a new resonance whose mass $(M_H)^{\rm exp} \sim$ 680(15) GeV is again in the expected mass range. As for the total width, the $\gamma\gamma$ data tend to place an upper limit $(\Gamma_H)^{\rm exp}<$ 50 GeV (consistently with the very loose indication $(\Gamma_H)^{\rm exp}=$ 29(20) GeV from the 4-lepton sample). 

Since the analogous CMS full distributions, for the high-mass 4-lepton and inclusive $\gamma\gamma$ events, are still missing, we first considered in Sect.5 previous CMS small-sized samples for these two channels. The lower statistics results do not contradict the existence of a new resonance in the relevant mass region but do not add much to the discussion, We thus considered two more recent CMS analyses which indicate a modest 1.6-sigma (local) excess at 675(25) GeV  in the $ b \bar b + \gamma\gamma$ final state and a sizeable 3.3-sigma (local) excess at 650(40) GeV in the invariant mass of $\gamma\gamma$ pairs produced in $pp$
double-diffractive scattering. Altogether, these two CMS results
point toward a new resonance with a mass of about 670(20) GeV.

In view of the good agreement with the indications extracted from the
ATLAS data, it is natural to wonder about the present, overall statistical
significance. To this end, one should first take into account that, when comparing with some theoretical prediction which refers to a definite mass region, as in the case of Eq.(\ref{prediction}),
local excesses should maintain intact their statistical significance
and {\it not} be downgraded by the so called look-elsewhere effect.

For this reason, since the correlation among the above measurements
is presumably very small and all effects are concentrated in the same mass region (the 2.5-sigma excess at
at 680(30) GeV followed by the 3-sigma defect at 740(30) GeV in the ATLAS 4-lepton channel, the 3.3-sigma excess at 684(8) GeV in
the ATLAS $\gamma\gamma$ channel, the 1.6-sigma excess at 675(25) GeV in the
CMS $ b \bar b + \gamma\gamma$ final state and the 3.3-sigma excess at 650(40)
GeV in the CMS diphoton events produced in $pp$ double-diffractive
scattering) one may observe that the cumulated statistical significance  has  reached a substantial level. Indeed, there are several, completely different analyses, done by two different experiments, that, in spite of not having been optimized for the kind of resonance we have predicted, show excesses of events exactly in the mass region of our interest. 

Still, there is no question, announcing a discovery would be too premature for, at least, two reasons.

First, we have only taken into account the intrinsic $\sqrt{N}$
statistical errors and neglected the systematic uncertainties which
are needed in an experimental analysis claiming that something new
has been discovered. In particular this remark is true for a typical
Higgs search at LHC which has thousands of nuisance parameters that
are determined together with the physical parameters of interest. 
As a definite example of systematic uncertainty, we have considered in Sect.5 the 3.3-sigma excess at 650(40) GeV in the invariant mass of diphoton events produced in $pp$ double-diffractive scattering, but exactly the same discussion could be repeated for the other effects. In the end, after a more careful re-evaluation of all backgrounds and by increasing the statistics,  
the cumulated significance could be considerably weakened with respect to the present sizeable level.  

Second, there are other final states which, at present, show no
appreciable difference from the estimated background. While this represents a general warning, it is not a good reason to ignore the indications we have brought to the attention of the reader. Actually, it is just the opposite because, given the present energy and luminosity of LHC, the hypothetical
second resonance is too heavy to be immediately seen in all possible
final states. Instead, the existence of deviations, in some channel
and in a particular energy range, should give motivations to sharpen
the analysis in the other sectors. 

With all possible caveats, we thus conclude, this intriguing situation could be definitely clarified when two crucial pieces of information which are still missing will be available: the final RUN2 high-mass distributions for the charged 4-lepton channel and for the inclusive $\gamma\gamma$ final state of the CMS Collaboration, possibly including this time a dedicated study to prove or disprove our prediction.

\end{document}